\documentclass[twocolumn,superscriptaddress]{revtex4-1}%
\usepackage{amsmath}%
\usepackage{amsfonts}%
\usepackage{amssymb}%

\usepackage{graphicx}

\usepackage{fullpage}

\usepackage{dsfont}
\usepackage{hyperref}
\usepackage{bm}
\usepackage{multirow}
\usepackage{color}
\usepackage{float}

\begin{document}

\title{Temporal evolution of low-temperature phonon sidebands in WSe$_2$ monolayers}

\author{Roberto Rosati}

\author{Samuel Brem}
\author{Ra\"ul Perea-Caus\'in}

\affiliation{Chalmers University of Technology, Department of Physics,
412 96 Gothenburg, Sweden}

\author{Koloman Wagner}
\author{Edith Wietek}
\author{Jonas Zipfel}

\affiliation{Department of Physics, University of Regensburg, Regensburg D-93053, Germany}

\author{Malte Selig}

\affiliation{Institute of Theoretical Physics, Technical University Berlin, 10623 Berlin, Germany}

\author{Takashi Taniguchi}

\affiliation{International Center for Materials Nanoarchitectonics,  National Institute for Materials Science, Tsukuba, Ibaraki 305-004, Japan}

\author{Kenji Watanabe}

\affiliation{Research Center for Functional Materials, National Institute for Materials Science, Tsukuba, Ibaraki 305-004, Japan}

\author{Andreas Knorr}

\affiliation{Institute of Theoretical Physics, Technical University Berlin, 10623 Berlin, Germany}

\author{Alexey Chernikov}
\affiliation{Department of Physics, University of Regensburg, Regensburg D-93053, Germany}
\author{Ermin Malic}

\affiliation{Chalmers University of Technology, Department of Physics,
412 96 Gothenburg, Sweden}

\begin{abstract}
Low-temperature photoluminescence (PL) of hBN-encapsulated monolayer tungsten diselenide (WSe$_2$) shows a multitude of sharp emission peaks below the bright exciton. Some of them have been recently identified as phonon sidebands of momentum-dark states. However, the exciton dynamics behind the emergence of these sidebands has not been revealed yet. In this joint theory-experiment study, we theoretically predict and experimentally observe time-resolved PL providing microscopic insights into thermalization of hot excitons formed after optical excitation. In good agreement between theory and experiment, we demonstrate a spectral red-shift of phonon sidebands on a timescale of tens of picoseconds reflecting the phonon-driven thermalization of hot excitons in momentum-dark states. Furthermore, we predict the emergence of a transient phonon sideband that  vanishes in the stationary PL.  
The obtained microscopic insights are applicable to a broad class of 2D materials with multiple exciton valleys.
\end{abstract}
\maketitle

Monolayer transition metal dichalcogenides (TMDs) have attracted large attention in current research due to their remarkable excitonic landscape 
including optically accessible bright excitons as well as spin- and momentum-dark states \cite{Wang18,Mueller18,Malic18,Mak10,Splendiani10,Xiao12,He14,Chernikov14,Zhang15}.
Furthermore, the excitonic properties become even richer when considering van der Waals heterostructures \cite{Novoselov16}
 with the emergence of spatially separated interlayer excitons \cite{Kunstmann18,Arora17,Miller17,Merkl19,Ovesen19} as well as twist-angle-dependent moir\'e exciton features \cite{Seyler19,Jin19,Alexeev19,Tran19,Brem20b}. 
At moderate and high temperatures the optical response in TMDs is dominated by the bright exciton. However, the latter is indirectly affected by the presence of dark excitonic states, 
which are most prominent in tungsten-based TMDs, e.g. via an increased linewidth broadening \cite{Selig16,Brem19}, reduced/enhanced temperature-dependent photoluminescence (PL) (reflecting the relative position of bright and dark excitonic states) \cite{Zhang15,Selig18} 
as well as altered diffusion in spatially-resolved experiments \cite{Cadiz18,Perea19,Cordovilla19,Zipfel20,Rosati20}.

\begin{figure}[t!]
\centering
\includegraphics[width=0.75\linewidth]{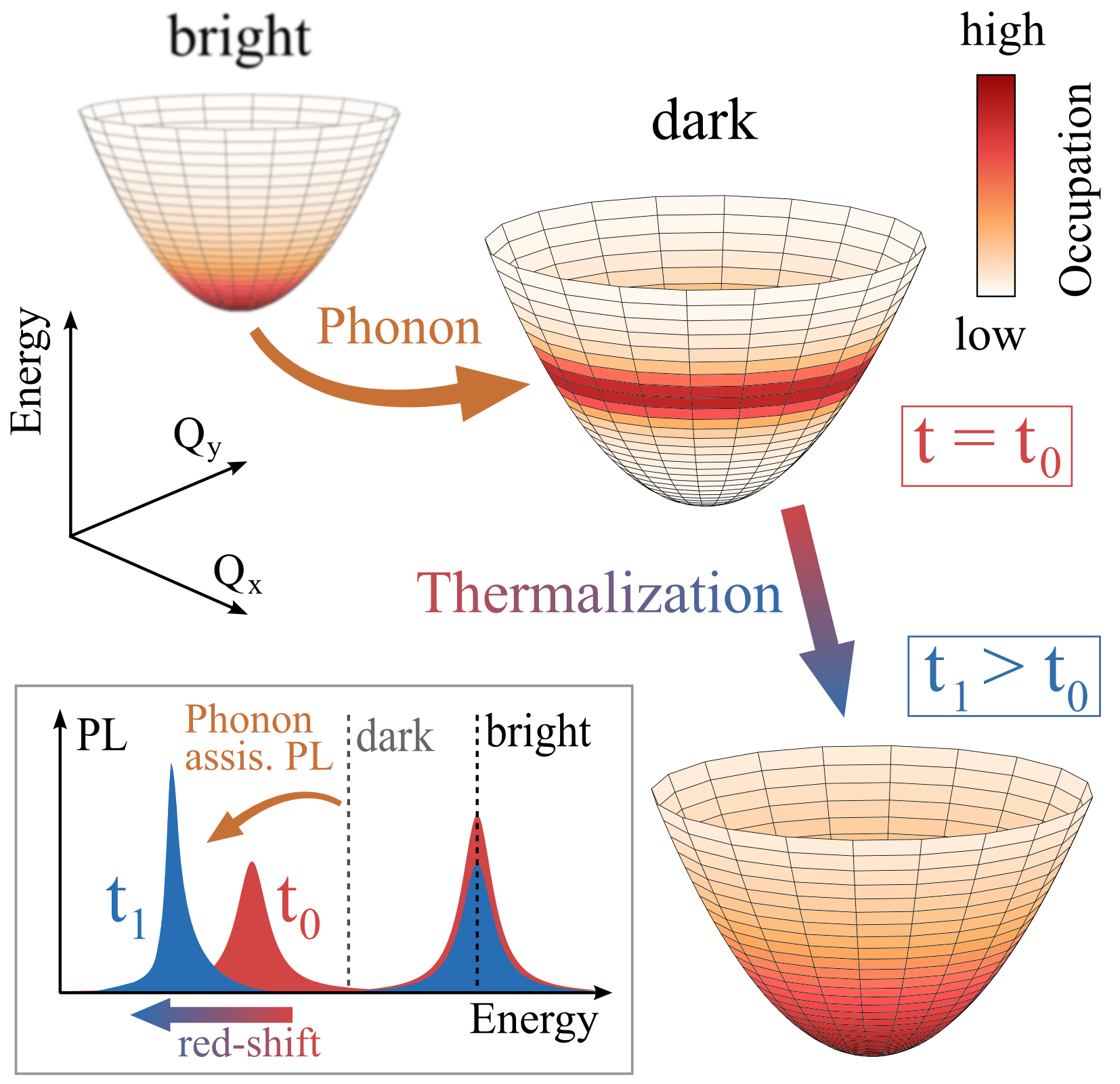}
\setlength{\belowcaptionskip}{-10pt}
\caption{\textbf{Sketch of phonon sideband evolution.} Phonon-driven scattering from the bright into the dark exciton results in a hot exciton distribution in the dark state. The latter can radiatively recombine via emission of a phonon giving rise to
a transient phonon sideband  at an early time $t_0$ (red peak in the inset).
Through thermalization of excitons, this sideband shifts to lower energies  (at a later time $t_1>t_0$) and forms a high-energy tail (blue peak) reflecting the equilibrium distribution of excitons. Thus, the red-shift of the phonon-sideband is a measure for the exciton thermalization in momentum-dark states. 
\label{fig:Fig1}}
\end{figure}

Cryogenic photoluminescence studies  of hBN-encapsulated tungsten-based TMDs, such as WSe$_2$, reveal a multitude of 
sharp emission peaks \cite{Courtade17,Ye18,Barbone18,Lindlau17,Li19,Liu19,Schneider20,Brem20}. Here, hBN-encapsulation gives rise to a strong decrease of the exciton linewidth \cite{Cadiz17b,Ajayi17} via a reduced dielectric disorder \cite{Raja19}.
While some of the observed low-energy peaks have been attributed to spin-dark excitons \cite{Li19,Liu19}, trions \cite{Wang14,Courtade17,Liu19a}, and biexcitons \cite{Ye18,Barbone18}, the origin of the peaks located at  50-80 meV below the bright exciton could be just recently ascribed to phonon-assisted recombination of momentum-dark  excitons \cite{Brem20,He20,Courtade17,Lindlau17,Ye18,Barbone18}.  
Importantly, phonon-assisted emission provides access to excitons with finite center-of-mass momenta to directly monitor their time-dependent distribution and study non-equilibrium relaxation dynamics \cite{Umlauff98}. 
Hot exciton distributions could play a crucial role for the optical response especially in tungsten-based TMDs, which exhibit a number of dark excitonic states located energetically below the bright exciton \cite{Malic18,Zhang15,Deilmann19}.

Here, we present a joint theory-experiment study on time-resolved photoluminescence focusing in particular on the temporally resolved formation of phonon sidebands in hBN-encapsulated WSe$_2$ monolayers. We combine time-resolved PL measurements with a microscopic theory that is based on exciton Wannier equation \cite{Koch06} and the generalized Elliot formula including phonon-assisted recombination channels \cite{Brem20}.
We theoretically predict and experimentally demonstrate transient low-energy PL signals  reflecting  the presence of hot excitons in momentum-dark states, cf. Fig. \ref{fig:Fig1}.
Due to a specific exciton landscape we also predict a transient peak stemming from K$\Lambda$ excitons.
As hot-exciton occupation thermalizes via phonon-induced  inter- and intravalley scattering, these transient features eventually vanish in the stationary PL reflecting the timescale of the respective scattering mechanisms.

\section{Results}
\subsection{Theoretical approach}
To investigate the dynamics of phonon-assisted PL from dark and bright excitons in the hBN-encapsulated WSe$_2$ monolayers, we need to reveal the exciton formation, thermalization and radiative recombination.
Considering the single-particle dispersion \cite{Kormanyos15} and solving the exciton Wannier equation \cite{Haug09,Selig16,Selig18,Brem18}, we obtain the exciton states
$\vert\alpha\rangle\equiv\vert\mathbf{Q},v\rangle$ characterized by the valley index $v$, the center-of-mass momentum $\mathbf{Q}$
and the exciton energy $E_\alpha=E_v+\hbar^2|\mathbf{Q}|^2/(2M_v)$ with $M_v$ as the total valley-dependent mass. Due to considerable energy separations, 
we restrict our attention to the $1s$ states of the bright excitons ($KK$) and the momentum-dark excitons (KK$^\prime$, $K\Lambda$) lying according to the solution of the Wannier equation approximately 46 and 36 meV below $KK$, respectively (cf. also Refs. \cite{Brem20,Deilmann19}). 
We assume the case of a spatially-uniform distribution and focus on the dynamics of incoherent exciton occupations
$N_{\alpha} =\left\langle \hat X^\dagger_{\alpha} X_{\alpha} \right\rangle $,
where $X^{(\dagger)}_{\alpha}$ are 
exciton annihilation (creation) operators for the state $\vert\alpha\rangle$ \cite{Katsch18}. 
Exploiting the Heisenberg equation and the many-particle Hamilton operator \cite{Kira06,Malic13,Haug09}, 
we derive an equation of motion for the exciton occupation \cite{Selig18,Merkl19}   in the low excitation regime
\begin{equation}\label{SBE}
\dot{N}_\alpha=\frac{1}{\hbar}\sum_{\alpha_B}(-2\gamma N_\alpha\delta_{\alpha,\alpha_B} + \Gamma_{\alpha\alpha_B}|p_{\alpha_B}|^2)+\!\!\left.\dot{N}_\alpha \right\vert_{\text{sc}}
\end{equation}
with $\alpha_B$ describing the set of bright states $\vert\alpha_B\rangle\equiv\vert \mathbf{Q}\approx 0,v=KK\rangle$.
The first term in equation (\ref{SBE}) takes into account the losses due to the direct radiative recombination with  $\gamma$  describing the radiative rate within the light cone  \cite{Selig16,Selig18,Brem18}. The
second term describes the formation of incoherent excitons due to phonon-driven transfer from 
the excitonic polarization $p_{\alpha_B}\equiv \left\langle X_{\alpha_B} \right\rangle$ that is often referred to in literature as coherent excitons \cite{Selig18}. The 
latter are optically excited in the light cone by an electromagnetic
field $\mathbf{A}(t)$ through
$\left.\dot{p}_{\mathbf{Q}}(t)\right|_{op}\propto \mathbf{M}\cdot\mathbf{A}(t) \delta_{\mathbf{Q}, 0}$ with $\mathbf{M}$ 
describing the excitonic optical matrix elements \cite{Selig18,Brem18}. 
The formation of incoherent excitons is driven by exciton-phonon scattering rates $\Gamma_{\alpha\alpha^\prime}$ 
describing scattering from state $\vert\alpha^\prime\rangle\equiv\vert  \mathbf{Q}^\prime v^\prime\rangle$ to $\vert\alpha\rangle\equiv\vert \mathbf{Q} v\rangle$  via emission or absorption of acoustic and optical phonons \cite{Selig18,Brem18}.
Finally, the last term in equation (\ref{SBE}) describes the scattering contribution giving rise to a thermalization of optically excited excitons. In this work we consider high-quality 
hBN-encapsulated TMD monolayer samples, where disorder and the associated scattering channels are strongly suppressed \cite{Cadiz17b,Ajayi17, Raja19}. 
We  restrict our attention to the low-excitation regime, where the main source of scattering
is given by exciton-phonon interactions. The scattering-induced dynamics can thus be written as
$\left.\dot{N}_\alpha\right|_{sc}=\Gamma^{\text{in}}_{\alpha}\!-\!\Gamma^{\text{out}}_{\alpha}N_{\alpha}$,
where the first (second) term describes the in- (out-) scattering dynamics of 
$N_{\alpha}$ with $\Gamma^{\text{in}}_{\alpha}\equiv\sum_{\alpha^\prime} \Gamma_{\alpha \alpha^\prime} N_{\alpha^\prime}$ and $\Gamma^{\text{out}}_{\alpha}=\sum_{\alpha^\prime}\Gamma_{\alpha^\prime\alpha}$. 

Once we know the temporal evolution of excitonic populations, we can also determine  the time-resolved photoluminescence  \cite{Brem20}  
\begin{equation}\label{PhaPL}
I(E,t)= \frac{2 |M|^2\left[
I_d(E,t)+I_{ind}(E,t)\right]}{(E_{0,KK}-E)^2+(\gamma+\Gamma_{0,KK})^2}
\end{equation}
including the direct ${I_d(E,t)=\gamma N_{0,KK}(t)}$ and phonon-assisted indirect PL ${I_{ind}(E,t)=\sum_{\alpha, \beta,\pm} \!\!|D^{(\beta)}_{\alpha}|^2 N_\alpha(t) \eta^{\beta \pm}_{\alpha} \frac{\Gamma_{\alpha}}{(E_{\alpha}\pm \epsilon_{\beta}-E)^2+\Gamma_\alpha^2}}$. 
Here, we have introduced the scattering rate  $\Gamma_{\alpha}=\Gamma^{\text{out}}_{\alpha}/2$ and the exciton-phonon matrix element $D^{(\beta)}_{\alpha}$ describing the scattering from the exciton state $\alpha$ to the virtual states in the light cone via emission/absorption of the phonon $\beta$ with the  energy $\epsilon_\beta$. The latter enters in $\eta^{\beta \pm}_{\alpha}=\tilde{N}_\beta +1/2\mp 1/2$ with the phonon occupation $\tilde{N}_\beta $, which is assumed to be well described by the Bose-Einstein distribution (bath approximation \cite{Malic13}). In contrast to previous studies, we go beyond the stationary PL solution (stemming from an equilibrium exciton distribution) and explicitly include  
time-dependent excitonic populations $N_\alpha(t)$ evolving according to equation (\ref{SBE}). This allows us to investigate the impact of hot excitons on the optical response. In particular, the formation of phonon sidebands crucially depends on the time-dependent population of the involved dark excitonic states. Thus, we expect transient features in the PL spectrum reflecting hot excitons and their thermalization towards the equilibrium distribution.

\subsection{Low-temperature exciton dynamics}

\begin{figure}[t]
\centering
\includegraphics[width=\linewidth]{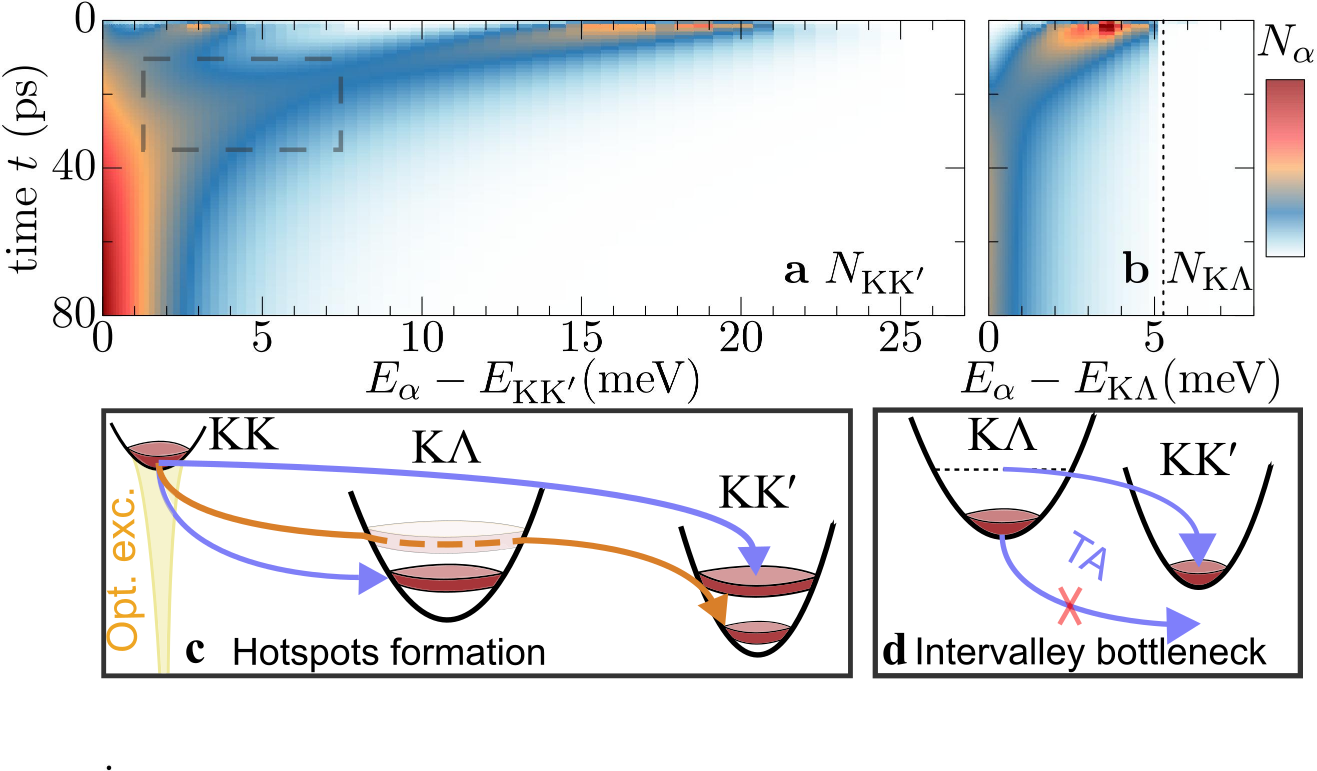}
\caption{\textbf{Exciton formation and thermalization.} Temporal evolution of the excitonic occupation $N_\alpha(t)$ for (\textbf{a}) KK$^\prime$  and (\textbf{b}) K$\Lambda$ excitons as a function of the excess energy $E_{\alpha}-E_v$. Sketches illustrating (\textbf{c}) exciton thermalization and formation of transient hotspots as well as (\textbf{d}) a phonon bottleneck for the intervalley thermalization after a few tens of ps.  The thin dashed line in (\textbf{b}) and (\textbf{d}) marks the minimal energy  for intervalley phonon emission. The high-energy hotspot in K$\Lambda$ has a very short lifetime, since excitons quickly scatter further down to KK$^\prime$ states.
\label{fig:Fig2}}
\end{figure}

Exploiting the generalized Elliot formula (\ref{PhaPL}) and the microscopic description of the  exciton dynamics  (\ref{SBE}) we have access to the time-resolved optical response of TMDs. 
Figure \ref{fig:Fig2} illustrates the temporal evolution of exciton occupations $N_\alpha(t)$ for the momentum-dark KK$^\prime$ and K$\Lambda$ states in hBN-encapsulated WSe$_2$ monolayers after an ultrafast optical excitation resonant to the bright exciton X$_0$, cf. Fig. \ref{fig:Fig2}a-b. 
The occupation of KK$^\prime$ excitons shows two hotspots: The hotspot with a higher energy is induced directly via emission of optical phonons from coherent excitons in the bright state 
(see second term in equation \ref{SBE} and Fig. \ref{fig:Fig2}c). The other hotspot with a smaller excess energy is induced by scattering from other incoherent states 
(see third term in equation \ref{SBE}), in particular via K$\Lambda$ excitons 
(e.g. through emission of acoustic M phonons).  In contrast, the occupation of K$\Lambda$ shows only one (longer lived) hotspot induced mostly by direct emission of higher-energy optical phonons from the bright exciton, cf. Fig. \ref{fig:Fig2}c. All these hot excitons  thermalize via intravalley scattering with acoustic phonons, thus showing a continuous decrease of the excess energy together with a broadening of the distribution. 

During this thermalization process the  intervalley scattering can redistribute excitonic population from one valley to another depending crucially on the relative spectral separation of the states.
 When the separation is large enough that phonon emission becomes possible, the intervalley scattering time decreases abruptly. This is the case for $K\Lambda$ excitons with excess energies larger than approximately 5 meV. The energy of these states is then larger than $E_{\text{KK$^\prime$}}+\epsilon_{\text{TA,M}}$ with $\epsilon_{\text{TA,M}}$ as the energy of acoustic intervalley phonons. The latter is indicated by the dashed line in Fig. \ref{fig:Fig2}b and illustrates the abrupt depopulation of  K$\Lambda$ states above this energy. However, at 20 K the stationary occupation (described by the Boltzmann distribution) of K$\Lambda$ excitons at an energy of 5 meV above $E_{\text{K}\Lambda}$ is very small. As a consequence, once the distribution of K$\Lambda$  excitons is thermalized, there is a phonon emission bottleneck for the states at the energy minimum of this valley, cf. Fig. \ref{fig:Fig2}d. Here,  the intervalley scattering to KK$^\prime$ states can only take place via relatively unlikely absorption of intervalley phonons at the considered low temperatures. If K$\Lambda$ is still overpopulated with respect  to KK$^\prime$ at this stage of the evolution, the population redistribution from K$\Lambda$ to KK$^\prime$  becomes very slow, which will turn out to give rise to some interesting
transient features in the PL spectrum. 
The states close to the minimum of $E_{\text{KK}^\prime}$ become effectively populated after a few picoseconds, when the first occupation hotspot thermalizes. The energetically  higher second hotspot  merges with the former during the thermalization process, thus creating a transient broadening of the energetic distribution (cf. dashed rectangle at approximately 20 ps in Fig. \ref{fig:Fig2}a).

\subsection{Time-dependent phonon sidebands}

\begin{figure}[t]
\centering
\includegraphics[width=0.85\linewidth]{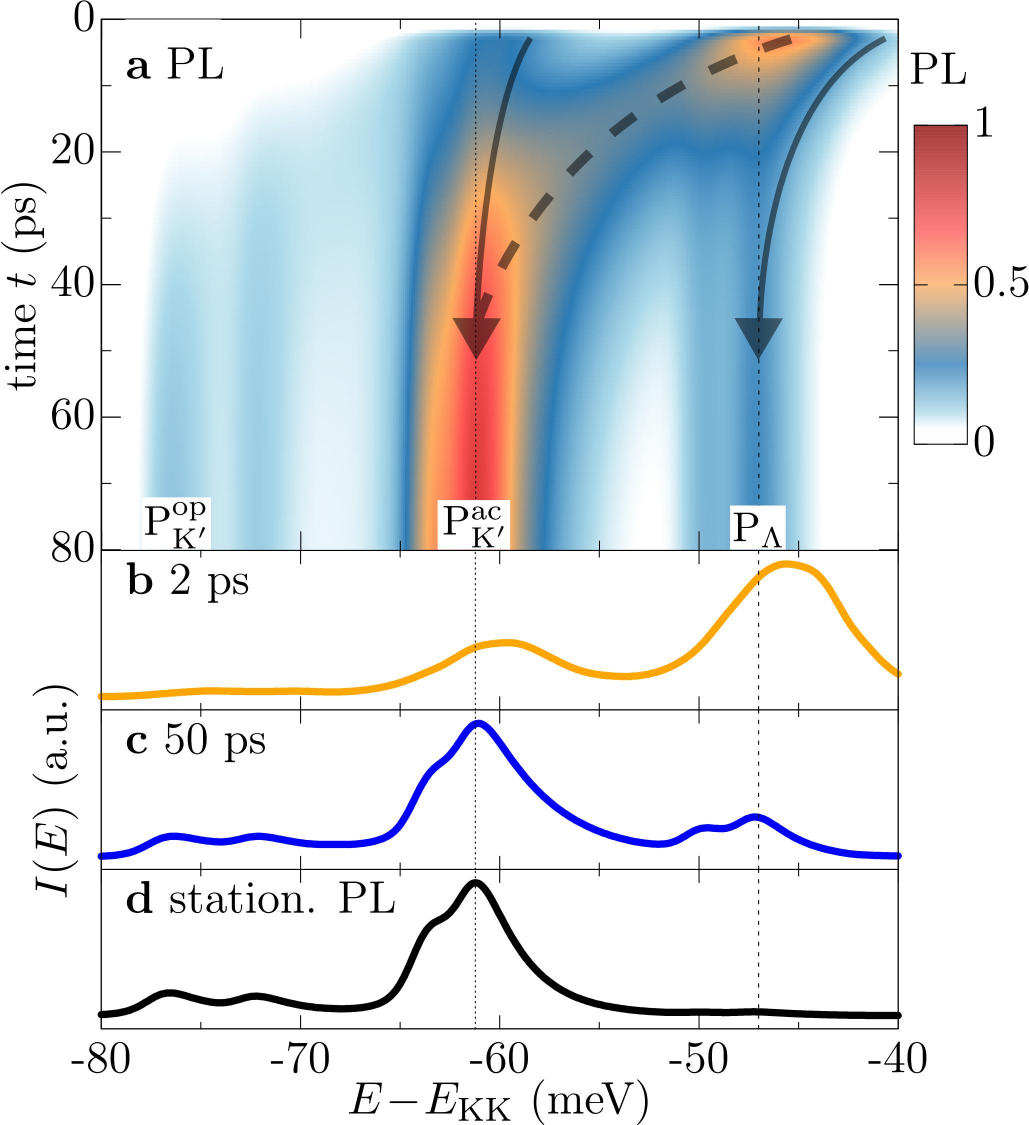}
\caption{\textbf{Low-temperature PL dynamics.}
(\textbf{a}) Time-resolved photoluminescence in hBN-embedded WSe$_2$
monolayer at 20 K focusing on the low-energy window, where phonon sidebands P$_\Lambda$ and P$_{\text{K}^\prime}$ appear. The PL has been normalized to the intensity of the bright exciton at each time.
We observe clear red-shifts of the phonon sidebands marked by arrows. Cuts at given times are given in (\textbf{b}) and (\textbf{c}) to better illustrate the transient character of P$_\Lambda$. (\textbf{d}) Stationary phonon sideband PL profile calculated assuming an equilibrium exciton distribution.  The observed sub-peak structure is due to the energy difference of the involved acoustic and optical phonon modes.
\label{fig:Fig3}}
\end{figure}

After having analyzed the low-temperature evolution of excitonic occupations, we can now understand the time-resolved photoluminescence. We focus on the contribution stemming from incoherent excitons, while strong signals from coherent excitons are expected in the sub-picosecond range \cite{Selig16}. 
Furthermore, we focus on low-temperature PL, where phonon sidebands below the bright exciton $X_0$ are known to strongly contribute to the luminescence \cite{Brem20}. Thus, Fig. \ref{fig:Fig3} illustrates only the spectral region below the bright exciton. While Fig. \ref{fig:Fig3}a shows the full time- and energy-resolved PL, in Figs. \ref{fig:Fig3}b-c we show cuts at two specific times (soon after optical excitation and at 50 ps). The exciton occupation in the momentum-dark  K$\Lambda$ and KK$^\prime$ states (Figs. \ref{fig:Fig2}a-b) induces the formation of the phonon sidebands  P$_\Lambda$ and P$_{\text{K}^\prime}$.  Having microscopic insight into the exciton dynamics, we can clearly ascribe these PL signals to the indirect radiative recombination induced by acoustic-TA phonon emission from the bottom of KK$^\prime$ and K$\Lambda$ valleys, respectively. 
Furthermore, we also observe additional lower-energy signals $P_{\text{K}^\prime}^{\text{op}}$ with a smaller intensity stemming from the emission of optical phonons from KK$^\prime$ excitons. 

We find that the P$_\Lambda$ phonon sideband is only a transient signal and almost vanishes in the stationary PL \cite{Brem20}, cf. Fig. \ref{fig:Fig3}d.
The stationary character is due to the low occupation of K$\Lambda$ excitons compared to
the energetically lower KK$^\prime$.
The reason for the transient phonon sideband is the intervalley phonon bottleneck between the 
 K$\Lambda$ and KK$^\prime$ excitons (cf. Fig. \ref{fig:Fig2}d) resulting in a long-lived overpopulation of the energetically higher K$\Lambda$ states. Note that the emergence of this feature is sensitive to the relative spectral separation of these momentum-dark exciton states. For $E_{\text{K}\Lambda}-E_{\text{KK}^\prime}\gtrapprox  \epsilon_{\text{TA,M}}$, the intervalley thermalization via emission of TA phonons would be  much more efficient resulting in a smaller occupation of K$\Lambda$ excitons already at early times and thus the P$_\Lambda$ sideband would disappear much quicker. Thus, the formation of the P$_\Lambda$ phonon sideband could be experimentally tuned by changing the spectral alignment of momentum-dark excitons, 
e.g. through the application of small values of strain \cite{Conley13,Steinhoff15,Niehues18,Khatibi18}. 

For both P$_\Lambda$ and P$_{\text{K}^\prime}$ phonon sidebands, we predict a clear red-shift (Fig. \ref{fig:Fig3}a) reflecting the thermalization of high-energy hotspot in the occupation of KK$^\prime$ and in K$\Lambda$ excitons (Figs. \ref{fig:Fig2}a-b).  Hot excitons cool down toward the bottom
of their respective exciton dispersions via intravalley exciton-phonon scattering and give rise to a spectral shift of phonon sidebands in the first few tens of picoseconds, cf.  Fig. \ref{fig:Fig3}b-c. We find red-shifts in the range of a few meV  toward P$_{\text{K}^\prime}$ and P$_\Lambda$ (solid arrows) induced by the thermalization of KK$^\prime$ and K$\Lambda$ excitons with a relatively small excess energy, cf. the low-energy hotspots in the KK$^\prime$ and K$\Lambda$ exciton occupation in Figs. \ref{fig:Fig2}a-b. In addition, we also find a larger shift of PL intensity from P$_\Lambda$ to P$_{\text{K}^\prime}$ (dashed arrow) stemming from the thermalization of the higher-energy hotspot in the KK$^\prime$ occupation.  
The temporal evolution of the predicted red-shift is a measure for the exciton thermalization, which is driven by intra- and intervalley exciton-phonon scattering. As a result, the formation of phonon sidebands is expected to show a strong
temperature dependence.

\subsection{Temperature-dependent thermalization}

\begin{figure}[t]
\centering
\includegraphics[width=\linewidth]{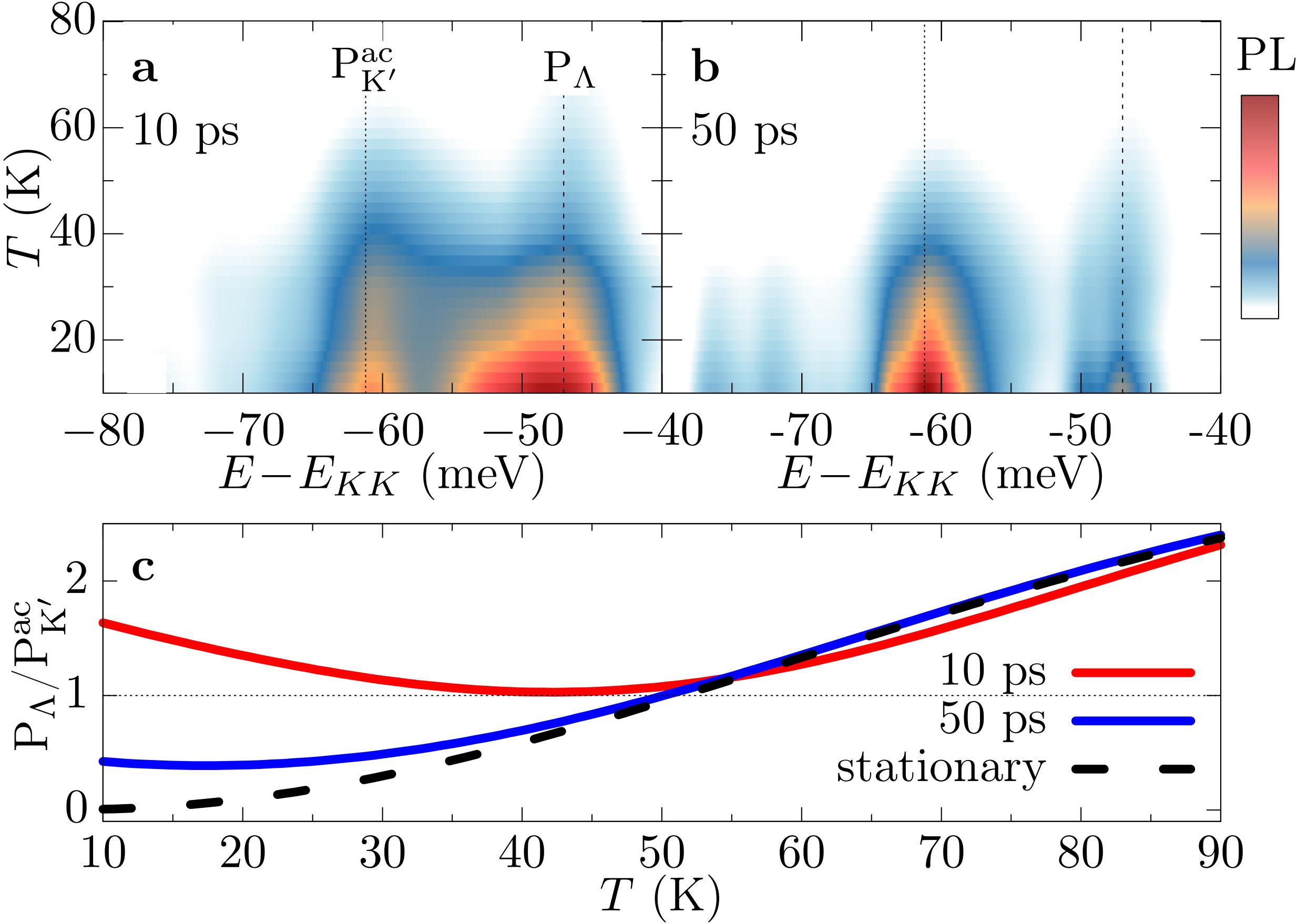}
\caption{\textbf{Temperature-dependent transient PL.}
Surface plot showing the PL as a function of temperature and energy
at (\textbf{a}) 10 ps and (\textbf{b}) 50 ps (normalized to the sideband/bright
exciton ratio at 10 K), respectively. (\textbf{c}) Temperature-dependent maximum intensity ratio of P$_\Lambda$ and P$_{\text{K}^\prime}^{\text{ac}}$ phonon sidebands. 
Hot excitons in KK$^\prime$ as well as temporary overpopulated $K\Lambda$ excitons give rise to  temperature-dependent transient PL signals before the stationary PL is reached. 
\label{fig:Fig4}}  
\end{figure}

Figure \ref{fig:Fig4} illustrates the impact of temperature on the temporal evolution of phonon sidebands. The temperature- and energy-dependent PL is shown at two fixed times of 10 ps and 50 ps.
We find that at both considered times the intensity of the phonon sidebands drastically decreases as the temperature increases.
In addition, at 10 ps we also observe  for temperatures below approx. 40 K a broad low-energy shoulder of P$_\Lambda$, which is not present at 50 ps. 
Although energetically closer to P$_\Lambda$, this optical response is related to P$_{\text{K}^\prime}^{\text{ac}}$ because it stems from the thermalization of the high-energy hotspot in the KK$^\prime$ exciton occupation, which at 10 ps has a low-energy tail energetically below the minimum of $K\Lambda$ (cf. Fig. \ref{fig:Fig2}a for 20 K). At low temperatures the exciton-phonon scattering is less efficient, thus requiring a longer time for the intravalley thermalization of hot excitons and resulting in low-temperature transient features in the PL spectrum. 
In contrast, at temperatures higher than 50K the faster exciton thermalization reduces drastically the weight of these  transient features associated with non-thermalized exciton occupations.

 Figure \ref{fig:Fig4}c illustrates the temperature-dependent intensity ratio between P$_\Lambda$ and P$_{\text{K}^\prime}^{\text{ac}}$ sidebands at two fixed times.  The dashed curve shows the stationary case for comparison, where exciton occupations correspond to the equilibrium Boltzmann distribution. 
 We find that at low temperature the ratio is nearly zero reflecting the extremely low occupation of K$\Lambda$ excitons resulting in a very low intensity of the P$_\Lambda$ phonon sideband. The higher the temperature, the larger is the occupation of K$\Lambda$ and the higher is the P$_\Lambda$/P$_{\text{K}^\prime}^{\text{ac}}$ ratio. For temperatures larger than approx. 50 K, the ratio even exceeds 1 despite the K$\Lambda$ exciton occupation (responsible for P$_\Lambda$) being still smaller 
than the KK$^\prime$ occupation. The reason lies in the larger spectral overlap of K$\Lambda$ excitons with the light cone in the bright exciton $X_0$, cf. the product of Lorentzian functions for indirect emission in \ref{PhaPL}.
 
 Now we consider two non-stationary situations: 
 At 10 ps, the intensity of the transient P$_\Lambda$ peak dominates even the low-temperature PL.
 At 50 ps, the P$_\Lambda$/P$_{\text{K}^\prime}^{\text{ac}}$ ratio has changed in favor of P$_{\text{K}^\prime}^{\text{ac}}$, however P$_\Lambda$ still keeps a noticeable height that is larger compared to the stationary PL. This reflects a transient overpopulation of $K\Lambda$ excitons with respect to the KK$^\prime$ states (Fig. \ref{fig:Fig2}d).  The transient P$_\Lambda$/P$_{\text{K}^\prime}^{\text{ac}}$ ratio at 50 ps already corresponds to the stationary case for temperatures larger than approximately 50 K, reflecting a completed intervalley thermalization between KK$^\prime$ and K$\Lambda$ excitons. 
While a finite signal from P$_\Lambda$ is expected at moderate temperatures already from the stationary regime \cite{Brem20} (see also black dashed line in Fig. \ref{fig:Fig4}c), the PL signal in P$_\Lambda$ at very low temperatures is a genuinely transient feature  originating from the slow intervalley exciton-phonon scattering and the resulting overpopulation of $K\Lambda$ excitons, as discussed above. Note that 
 in addition to exciton-phonon coupling, elastic scattering with defects can also  contribute to an intervalley thermalization. 
The strength of exciton-disorder scattering can be roughly estimated through the temperature-independent  linewidth. 
The latter is very small in hBN-encapsulated samples implying that this coupling mechanisms might not be crucial \cite{Cadiz17b,Ajayi17}. 

 \subsection{Theory-experiment comparison}

\begin{figure}[t]
\centering
\includegraphics[width=\linewidth]{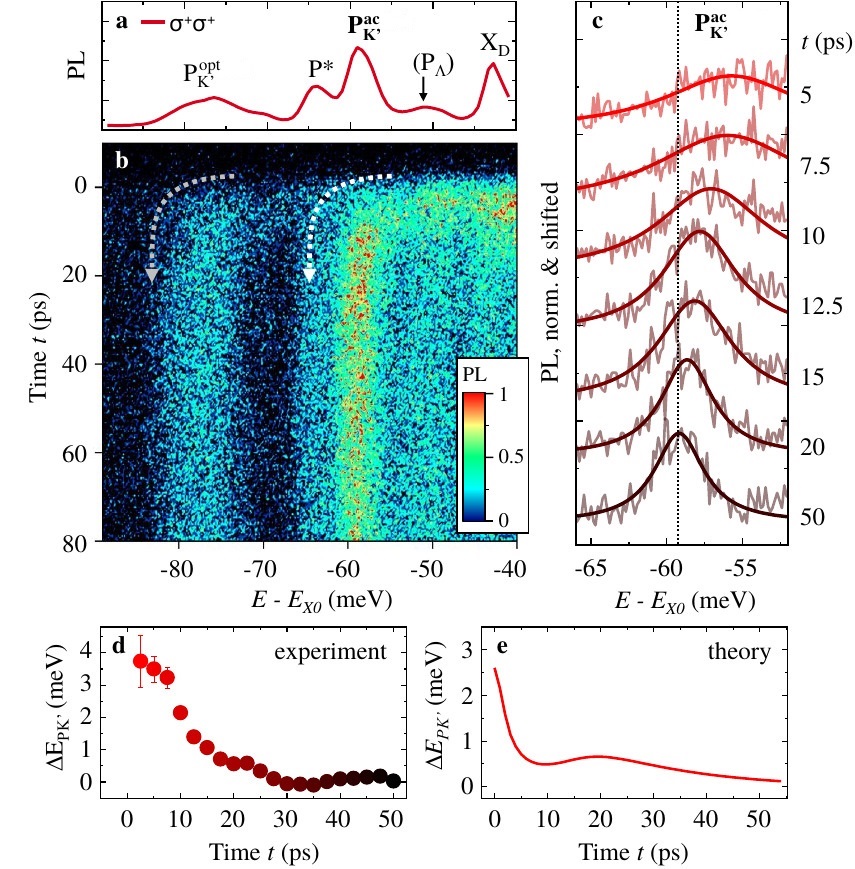}
\caption{\textbf{Experimental time-resolved PL spectra.}
(\textbf{a}) Time-integrated circularly co-polarized PL spectrum ($\sigma^+$ polarization both in excitation and detection) of a hBN encapsulated WSe$_2$ monolayer at 5K for pulsed excitation in resonance with the A-exciton transition.
The spectrum is presented in the range of the phonon-sidebands from dark states as function of energy relative to the A-exciton.
(\textbf{b}) Corresponding time-resolved streak camera image with the PL intensity plotted on a false-color scale.
Dashed lines are guides to the eye.
(\textbf{c}) Spectral cuts at selected times after the excitation in the spectral range of the P$^{\text{ac}}_{K'}$ phonon sideband together with phenomenological fit curves. 
(\textbf{d}) Time-resolved spectral shift of the P$^{\text{ac}}_{K'}$ peak energy extracted from the measurements in comparison to the theoretical prediction at 20 K presented in (\textbf{e}).
 \label{fig:Fig5}}
\end{figure}

To illustrate that the sideband features of non-equilibrium exciton populations also appear in experiments, we now discuss time- and spectrally-resolved photoluminescence measurements in hBN-encapsulated WSe$_2$ monolayers.
To match the scenario presented in the theory part we tune the excitation photon energy into resonance with the fundamental A-exciton transition (X$_0$) to create excitons within the light cone as well as employ co-circularly polarized injection and detection.
Similarly to the theoretical calculations, we focus on the emission at lower energies, where phonon sidebands of momentum-dark excitons appear, as illustrated in the time-integrated spectrum in Fig. \ref{fig:Fig5}a.
The corresponding time- and spectrally-resolved streak camera image of the PL is presented in Fig. \ref{fig:Fig5}b.

The data shows a pronounced peak about 40\,meV below $X_0$ stemming from z-polarized spin-dark exciton emission ($X_D$) detected due to the finite aperture in our measurements \cite{Wang17}.
At lower energies we observe characteristic, prominent peaks at approximately -60 and -80 meV, associated with PL from momentum-dark KK$^\prime$ excitons, cf. Fig. \ref{fig:Fig3}, under simultaneous emission of zone-edge acoustic (P$^{\text{ac}}_{K'}$) and  optical phonons (P$^{\text{op}}_{K'}$), respectively\,\cite{Brem20}.
Importantly, both (P$^{\text{ac}}_{K'}$) and (P$^{\text{op}}_{K'}$) exhibit very similar dynamics after the excitation with considerable changes in the lineshape and peak position during the first tens of picoseconds, closely resembling theoretical calculations presented in Fig. \ref{fig:Fig3}.

In addition, we consistently find a weak emission feature at about -50\,meV that could be assigned to the theoretically predicted transient sideband stemming from  K$\Lambda$ excitons (P$_{\Lambda}$), cf. Figs. \ref{fig:Fig3} and \ref{fig:Fig4}.
We note, however, that due to finite excitation density and weak unintentional doping of the sample we can not unambiguously exclude possible contributions from charged biexcitons that would emit at similar energies \,\cite{Ye18}.
Thus, while the emission dynamics in this spectral range appear qualitatively very similar to those theoretically predicted (compare Figs. \ref{fig:Fig5}b and \ref{fig:Fig3}), this consideration should limit a more detailed analysis.
Finally, the peak on the low energy shoulder of P$^{\text{ac}}_{K'}$, denoted as P* was recently demonstrated to stem from KK spin-dark excitons (X$_D$) \cite{Li19,Liu19} under the emission of a $\Gamma$-point optical phonon with specific symmetry allowing for spin-mixing of the conduction band states.
It is not included in the calculations presented above to keep the focus on the main features that appear under spin conservation.

For a more detailed analysis measured spectra of the P$^{\text{ac}}_{K'}$ sideband are presented in Fig. \ref{fig:Fig5}c at different times after the optical excitation.
As time passes by, the peak becomes considerably narrower and exhibits a shift to lower energies, reflecting the thermalization of hot excitons to the energetically lowest KK$^\prime$ states (Fig. \ref{fig:Fig2}) via interaction with intravalley acoustic phonons. 
In particular, this leads to a red-shift and reduced linewidth of the PL as the exciton distribution becomes more and more centered around the band minimum. 
Fig. \ref{fig:Fig5}d illustrates the temporal evolution of the relative energy shift that is in the range of a few meV and decreases to an equilibrium value on a timescale of several tens of picoseconds.
Figure \ref{fig:Fig5}e shows the corresponding theoretically calculated shift of the P$^{\text{ac}}_{K'}$ phonon sideband that is extracted from Fig. \ref{fig:Fig3}, illustrating a good qualitative agreement between theory and experiment. 
Both the magnitude of the absolute values and the equilibration time-scale from intravalley thermalization are  similar.
Interestingly, the energy shift in both theory and experiment also exhibit a small deviation from the continuous decrease at about 20 ps. 
In our calculations this effect stems from merging of the two hotspots in KK$^\prime$ exciton occupation (Fig. \ref{fig:Fig2}a) during their phonon-driven thermalization toward the energy minimum. 
Altogether, the overall consistency of the measured phonon sideband dynamics with theoretical predictions highlights the possibility to directly access non-equilibrium exciton populations of the dark states.

\section{Discussion}

In conclusion, the presented joint theory-experiment study provides a microscopic access to the formation dynamics of phonon sidebands in low-temperature photoluminescence spectra of hBN-encapsulated WSe$_2$ monolayers. 
We find both in theory and experiment the emergence of pronounced sidebands stemming from emission of acoustic and optical phonons. We track the temporal evolution of these sidebands, 
which are induced by phonon-assisted indirect emission from initially hot excitons in momentum-dark K$\Lambda$ and KK$^\prime$ states states. These transient features shift to lower energies reflecting the thermalization of hot excitons toward the energy minimum in the corresponding exciton bands. 
Due to a specific exciton landscape, we also find a genuinely  transient peak stemming from K$\Lambda$ excitons that  vanishes in the stationary photoluminescence.
The study of the formation dynamics of phonon sidebands sheds light on the underlying intra- and intervalley exciton thermalization and in particular on the phonon-driven  thermalization between momentum-dark excitonic states.

\section{Methods}

\subsection{Theoretical methods}
The investigated exciton thermalization is driven by exciton-phonon scattering that has been treated in the second-order Born-Markov approximation \cite{Selig18,Brem18}. 
As phonon modes we take into account  longitudinal and transverse acoustic (LA, TA) and optical (LO, TO) modes as well as  
the out-of-plane A$_1$ optical mode, which provide the most efficient scattering channels 
\cite{Jin14}. Phonon energies are extracted from density functional
theory studies \cite{Jin14} and summarized in 
Table \ref{tab:phonon}.
\begin{table}[h]
\centering
\begin{tabular}{|c|c|c|c|c|}
  \hline \hline
  \multicolumn{5}{|c|}{Phonon energies (meV)} \\
    \hline
  Mode &  $\Gamma$ & K & M & $\Lambda$ \\%
	\hline%
    TA/LA & 0./0. & 15.6/18.0 & 15.3/16.3 & 11.6/14.3 \\
    TO/LO(E')& 30.5/30.8 & 26.7/31.5 & 28.4/31.8 & 27.3/32.5 \\
    A$_1$ & 30.5 & 31.0 & 29.8 & 30.4 \\
 \hline \hline
\end{tabular}
\caption{Relevant phonon energies; for $\Gamma$ acoustic modes a linear dispersion with sound velocity $v=3.3 \times 10^5$cm/s is considered. 
Values are taken from DFT calculations \cite{Jin14}.}
\label{tab:phonon}
\end{table}  

For each intra- as well as intervalley scattering mechanism, the same deformation potentials have been used for the two acoustic and the three optical ones. 
Other studies suggested the possibility of a less-symmetric behaviour of exciton-phonon scattering \cite {Kaasbjerg12,Sohier18}, 
e.g. with a slightly larger contributions from TA phonons for  KK$^\prime$-KK intervalley scattering \cite{Sohier18}. 
This  could justify the less-evident sub-peak structures in experimental PL.
The required parameters for band-gap energy, spin–orbit splitting and effective masses  were taken from DFT calculations \cite{Kormanyos15}. 
For the radiative recombination rate $\gamma$ we obtain the value of 1.25 meV \cite{Brem19}, where we neglect its momentum-dependence justified by the small momenta involved 
in the bright states.  For the optical excitation, we consider a spatially-homogeneous pulse resonant to X$_0$ and exhibiting an amplitude FWHM of 0.2 ps.

\subsection{Experimental methods}

All experiments were performed on a WSe$_2$ monolayer encapsulated in high quality hexagonal boron nitride (hBN), deposited on a SiO$_2$/Si substrate. 
The individual layers were fabricated by micro-mechanical exfoliation of WSe$_2$ (from “HQgraphene”) and hBN (from NIMS) bulk crystals and subsequently stacked on top of each other by all-dry viscoelastic stamping onto a preheated (100$^\circ$C) Si/SiO$_2$ substrate. 
In order to enhance interlayer contact and avoid accumulation of impurities between the layers, the sample stack was annealed in high vacuum at 150$^\circ$C for 3 to 4 hours after each successful stamping step. 
The samples were characterized by linear reflectance and luminescence spectroscopy at cryogenic temperatures to confirm spectrally narrow linewidth and thus suppression of inhomogeneous broadening.

Time-resolved photoluminescence (PL) measurements were carried out in an optical microscopy cryostat in high-vacuum conditions at the temperature of 5\,K. 
The sample was excited by a 100\,fs pulsed Ti:sapphire laser with a repetition rate of 80\,MHz. 
Using second harmonic generation the photon energy was tuned into resonance with the A-exciton transition energy at 1.728\,eV. 
The excitation beam was circularly polarized by a quarter-wave plate and focused to a spot with a full-width-at-half-maximum of about 4\,$\mu$m (a larger spot than usual was chosen to keep the pump density reasonably low). 
The excitation power was set to 1\,$\mu$W. 
Assuming an effective absorption of about 30\% at the pump energy, estimated from absorption-type reflectance measurements as well as the overlap of excitation and absorption peak, the effective injected electron hole pair density is estimated to about 5$\times$10$^{10}$\,cm$^{-2}$.
The PL signal was collected in co-circular polarization and spectrally dispersed by a 300\,gr/mm grating. 
Steady state spectra were recorded by a cooled charged-coupled device camera, while time-resolved spectra were obtained using a streak camera, operating with a time resolution of several ps. 
For time-resolved measurements, the bright signal from the A-exciton resonance and excitation laser scattering were blocked by a spectral filter.
The time-zero was defined as half-rise time of the trion resonance.

\section{Acknowledgements}

This project has received funding from the Swedish Research
Council (VR, project number 2018-00734) and the European
Union’s Horizon 2020 research and innovation programme
under grant agreement no. 881603 (Graphene Flagship). The authors thank Alexander H\"ogele and Mikhail Glazov for fruitful discussions and Jonas D. Ziegler for the fabrication of the samples.
Financial support by the DFG via SPP2196 Priority Program (CH 1672/3-1) and Emmy Noether Initiative (CH 1672/1-1) is gratefully acknowledged.
The TUB group was funded by the Deutsche Forschungsgemeinschaft via projects 182087777 in SFB 951 (project B12, M.S., and A.K.).
K.W. and T.T. acknowledge support from the Elemental Strategy Initiative
conducted by the MEXT, Japan, Grant Number JPMXP0112101001,  JSPS
KAKENHI Grant Numbers JP20H00354 and the CREST(JPMJCR15F3), JST.



\end{document}